\documentclass[twocolumn,english,aps,prb,showpacs,floatfix]{revtex4}
\usepackage[T1]{fontenc}
\usepackage[latin1]{inputenc}
\usepackage{color}
\usepackage{graphicx}
\usepackage{amssymb}
\makeatletter

\usepackage{epsfig,amsmath,amssymb,color}
\usepackage{babel}
\makeatother
\begin{document}

\title{Static Holes in the Geometrically Frustrated Bow Tie Ladder}
\author{George B. Martins}
\affiliation{Department of Physics, Oakland University, Rochester, MI 48309, U.S.A.}
\author{Wolfram Brenig}
\affiliation{Institut für Theoretische Physik, Technische Universität Braunschweig,
D-38106 Braunschweig, Germany}
\date{\today{}}

\begin{abstract}
We investigate the doping of a geometrically frustrated spin ladder with static
holes by a complementary approach using exact diagonalization
and quantum dimers. Results for thermodynamic properties, the singlet density
of states, the hole-binding energy and the spin correlations will be presented.
For the undoped systems the ground state is non-degenerate, with translationally invariant nearest-neighbor spin correlations. In the
doped case,
we find that static holes polarize their vicinity by a localization of singlets
in order to reduce the frustration. This polarization induces short range
repulsive forces between two holes and an oscillatory behavior of the long
range two-hole energy.
For most quantities investigated, we find very good
agreement between the quantum dimer approach and the results from exact
diagonalization.
\end{abstract}

\pacs{75.10.Jm, 75.30.Hx, 75.40.Mg}

\maketitle

\section{Introduction}

Geometric frustration is a key factor, leading to exotic
phases in quantum spin systems.
Valence bond (VB) ordering, with and without breaking
of discrete lattice symmetries, is one possible type of
ground state symmetry. Examples are the spin-1/2 zig-zag
ladder \cite{Majumdar1969ab}, the checkerboard lattice
\cite{Brenig2002c,Fouet2003a}, the $j_{1}$-$j_{2}$-$j_{3}$ model
\cite{Singh1999a,Mambrini2006a}, or the Shastry-Sutherland model
\cite{Shastry1981a}. Another exotic phase is the spin liquid (SL)
which displays no apparent magnetic order but may break topological
symmetries, which presumably is the case for the two-dimensional (2D)
spin-1/2 kagom\'e antiferromagnet (AFM)
\cite{Zeng1990a,Waldtmann1998a,Misguich2005}.

Static holes, i.e. nonmagnetic defects, are an important probe of
such quantum phases. In case of enhanced VB correlations or VB order
without broken lattice symmetries, the situation is similar to spin-ladders
and the 2D Heisenberg AFM, where static holes generate
spin-1/2 moments in their vicinity \cite{Martins1996a,Laukamp1998a,Wessel2001}.
The latter induce triplet excitations at low-energies. In a SL, e.g.
in the kagom\'e AFM, a very different scenario has been observed,
in which singlet pairs accumulate close to the holes, rather than spin-1/2
moments \cite{Dommange2003a}. In case of VB states with spontaneous
breaking of the lattice symmetry \cite{Normand2002a}, and other valence-bond
ground states, \cite{Kolezhuk1998a} bound spin-1/2 moments also seem
to be absent. Apart from the properties of a single static hole, correlation
effects between two holes may shed light on the (de)confinement of
mobile carriers in frustrated quantum magnets \cite{Read1991a,Senthil2004a,Pollmann2006,Poilblanc2006a}
- at least for kinetic energies small compared to the exchange coupling,
i.e., $t\ll J$. Along this line, hole-hole interactions have been studied
recently in the kagom\'e AFM \cite{Dommange2003a} and in quantum
dimer models \cite{Misguich2004}. In both cases, deconfinement of
the static holes has been found. For two recent $\mu$SR-studies
of doped kagom\'e AFMs, see refs. \onlinecite{Fukaya2003a,Bono2004a}.

Additional insight into the response of frustrated spin models to
static holes can be obtained from a reduced description of the
singlet sector in terms of short-range resonating valence-bonds
\cite{Anderson1987a}, so-called quantum dimers
(QDs) \cite{Rokhsar1988a,Misguich2005}.
Classical dimer models on bipartite lattices allow for a height
representation
and tend to be confining, with a linearly or logarithmically attractive
`string potential' between holes \cite{Fisher1963a,Wu2003a}. Non-bipartite
lattices lack a height representation and allow for both, confining
as well as deconfining \cite{Misguich2002a,Fendley2002a,Misguich2003a,Krauth2003a,Huse2003a}
phases. Similar behavior is likely for QD models.

\begin{figure}[t]
\begin{center}\includegraphics[%
  width=0.6\columnwidth,
  keepaspectratio]{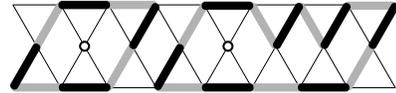}\end{center}
\vspace{-0.4cm}

\caption{\label{fig1} 1D Bow-Tie lattice structure of corner sharing
triangles. Spin-1/2 moments reside on all vertices. Thick grey (black)
bonds correspond to one QD configuration. Combination of grey and
black QDs represent one transition graph. Circles at the center correspond
to two static holes. }\vspace{-0.3cm}

\end{figure}

In this work we intend to shed further light on the role of static
holes in geometrically frustrated magnets by studying the spin-1/2 AFM
on a bow-tie ladder (BTL). The BTL is a one-dimensional
array of corner sharing triangles, as depicted in Fig. \ref{fig1}, and
may be viewed as the medial lattice of the two-leg ladder \cite{Misguich2003a}.
The AFM Heisenberg model on the BTL with nearest-neighbor exchange
\begin{equation}
H=\sum_{\left\langle lm\right\rangle }\mathbf{S}_{l}\cdot\mathbf{S}_{m}
\label{eq:1}\end{equation}
has been investigated extensively by linear spin wave theory, exact
diagonalization (ED) and density matrix renormalization group (DMRG)
in ref. \onlinecite{Waldtmann2000a}. This analysis has uncovered several
issues which remain open. In particular, ED on
up to $N=30$ sites suggests that
the BTL is qualitatively similar to the Kagom\'e AFM,
i.e. the system is a SL with a spin gap and a large number of singlets
($\propto N^{2}$) below the first triplet. However, ED does not allow
extrapolation of the triplet gap to the thermodynamic limit. DMRG
up to $N=120$ indicates a vanishing spin gap for $N\rightarrow\infty$.
Therefore, apart from a SL state it remains possible that the BTL
is critical or displays VB order with a very large unit cell.

Here, we will not focus on these properties of the clean BTL in the thermodynamic
limit. Instead, we will investigate the local correlations of \emph{two
static holes} - holons hereafter - introduced into finite BTLs. To this
end, results from two complementary methods will be discussed, namely, 
complete ED on the one hand, and restricted diagonalization in the
QD subspace on the other hand. Recently, QDs on a BTL lattice have
been considered, for the so-called $\mu$-model of shortest length resonance-moves
for the QDs. There, for certain \emph{ad-hoc} resonance amplitudes the QD
model was shown to be identical to the transverse-field Ising chain at
criticality
\cite{Misguich2003a}. In contrast to that study, we will account for
all resonance moves resulting from eqn. (\ref{eq:1}).

\section{Thermodynamic Properties}

\begin{figure}[t]
\begin{center}\includegraphics[%
  width=0.8\columnwidth,
  keepaspectratio]{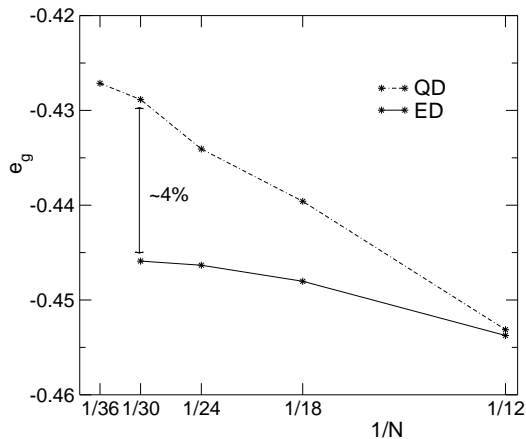}\end{center}
\vspace{-0.3cm}

\caption{\label{fig2}Ground state energy versus inverse system size for $N=12-30$
(ED) and $N=12-36$ (QD).}\vspace{-0.3cm}

\end{figure}
First, we briefly consider the clean system and compare the ground
state energy and specific heat, obtained by ED, with results from the QD approximation.
The generalized eigenvalue problem in the QD Hilbert space
uses the transition graphs depicted in Fig. \ref{fig1} and is set
up according to ref. \onlinecite{Zeng1995a} (to which we refer the reader
for details). Periodic boundary conditions (PBC) apply to all results
discussed in the following. For all systems studied we find the undoped
ground state to be non-degenerate, with translationally invariant
nearest-neighbor spin correlations, which is consistent with a spin
liquid state or a valence bond crystal with a unit cell large compared
to the system's sizes. In Fig. \ref{fig2}, we show the ground
state energy $e_{g}$ versus the inverse system size $1/N$. For QDs, proper
$N$ needs to be a multiple of 6. From Fig. \ref{fig2}, a deviation
of approximately 4-5\% between ED and QD can be extrapolated in the
thermodynamic limit. This has to be contrasted against the relative
dimension $D$ of the complete Hilbert space, $D_{0}$ of the singlet
sector, and $D_{QD}$ of the QD space, which are $D=2^{N}$, $D_{0}=N!/[(N/2)!(1+N/2)!]$,
and $D_{QD}=2^{N/3+1}$. For $N$=30, this implies
$D/D_{0}/D_{QD}\approx1.1\,10^{9}/(9.7\,10^{6})/(2.0\,10^{3})$.
In view of these ratios, the results in Fig. \ref{fig2} are in reasonable
agreement. The figure shows a very small `even/odd' oscillation with respect to 
$N/6$ in the QD results.

\begin{figure}[t]
\begin{center}\includegraphics[%
  width=0.8\columnwidth,
  keepaspectratio]{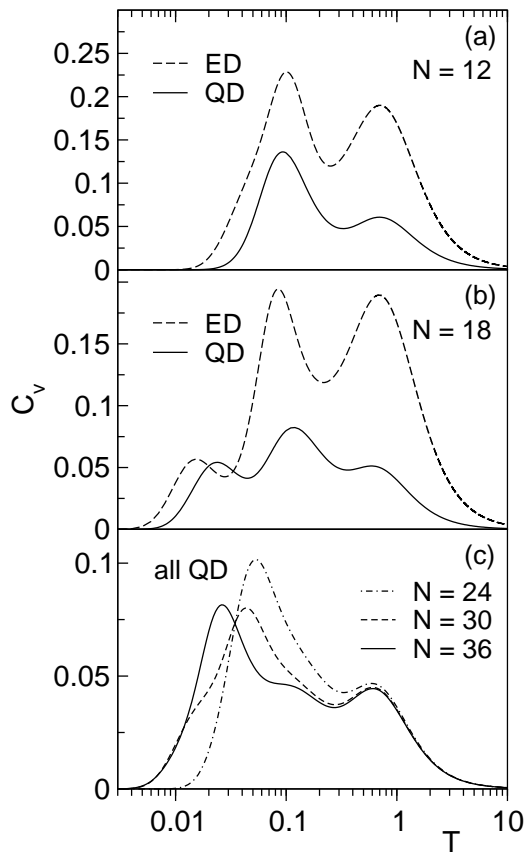}\end{center}
\vspace{-0.7cm}

\caption{\label{fig3}Specific heat $C_{V}$ versus temperature $T$. (a)
and (b): comparison of ED and QD for $N=12$ and $18$. (c) QD for
$N=24$-$36$.}\vspace{-0.3cm}

\end{figure}

A fingerprint of SL states in strongly frustrated magnets is the accumulation
of singlet states at low energies. In the kagom\'e AFM, the numer
of singlets below the first triplet has been observed to grow exponentially
with system size \cite{Waldtmann1998a}. In the BTL, power-law behavior
with an exponent $2$ has been suggested \cite{Waldtmann2000a}. The
singlet accumulation at low-energies leads to a characteristic singlet-triplet
double-peak structure in the specific heat $C_{V}$ \cite{Elser1989a,Zeng1990a,Waldtmann2000a,Ramirez2000a}.
For the kagom\'e AFM a substantial fraction of the singlet-peak in
$C_{V}$ is believed to be due to QD fluctuations \cite{Zeng1995a}.
For the BTL, this is an open issue. Therefore, in Fig. \ref{fig3}
we consider $C_{V}$ from QD and ED calculations. The latter agree
with identical results from ref. \onlinecite{Waldtmann2000a}. In view of
large finite-size effects our emphasis is on a comparison at identical
system sizes, rather than in the thermodynamic limit. First, there
is a remarkable similarity between the positions of peaks in ED and
QD in Fig. \ref{fig3}(a,b). Second, for the leftmost
peak in Fig. \ref{fig3}(b), at low temperatures, the magnitude of $C_{V}$
and the entropy
from ED and QD are comparable. I.e., we conclude that the low-$T$
peak in $C_{V}$ of the BTL is caused primarily by QD
fluctuations. As can be seen from Fig. \ref{fig3}(c), the
low-$T$ peak stabilizes as $N$ increases.
Third, in the high-energy region $T\gtrsim0.5$
a single peak can be observed which contains most of the entropy (note
the log $T$-scale). This entropy is significantly larger in the ED as
compared to the QD. This difference is due to triplet excitations
not present in the QD space. Fig. \ref{fig3}(c) shows that $C_{V}$
from QD has converged to the thermodynamic limit for $T\gtrsim0.1$
at $N>36$ - a system size we cannot reach with ED. Finally, the intermediate-$T$
feature in ED seems strongly system-size dependent, as already noted
in ref. \onlinecite{Waldtmann2000a}.

\section{Static Holes}

In this section, we consider the change in the singlet density-of-states
(DOS), the binding energy, and the spin correlations which stem from
introducing two holons into the BTL.

\begin{figure}[t]
\begin{center}\includegraphics[%
  width=0.78\columnwidth,
  keepaspectratio,
  angle=0]{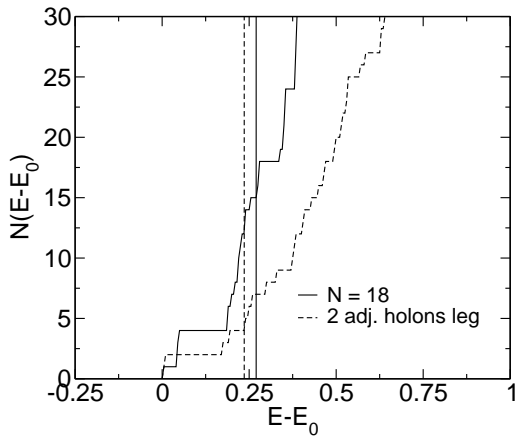}\end{center}
\vspace{-0.4cm}

\caption{\label{fig4}Number of singlets in ED
versus energy at $N=18$, contrasting $0$ holons (solid line) with 
$2$ holons (dashed). Solid (dashed) vertical line: triplet gap
for $0$ ($2$) holons. Holons are located at two adjacent sites on one
BTL leg. $E_{0}$ refers to ground state energy.}
\vspace{-0.4cm}
\end{figure}

\subsection{Density of States\label{sub:Density-States}}

Fig. \ref{fig4} shows the integrated singlet DOS versus energy, contrasting
the case of zero holons against that of two holons placed at adjacent
sites on the leg of a BTL with an even number of sites. The
main point of this figure is the number of singlets below the first
triplet. Evidently this number is reduced by introducing the holons.
A similar effect is observed for any other relative separation
of the holons and
also in the case of a single holon. This low-energy suppression of
singlet DOS originates from the reduction of frustration at the neighboring
sites of the holons, which favors binding of singlets in the vicinity
of the holons (see also section \ref{sub:Spin-Spin-Correlations}).
Consequently the low-energy singlet DOS decreases, while the triplet
gap is set by excitations distant from the holons and therefore less
affected. While similar effects have been reported for the kagom\'e AFM \cite{Dommange2003a}, this behavior is drastically different from VB states on
bipartite lattices, where static holes tend to bind spin-1/2 moments
which leads to an increase of low-energy triplet DOS \cite{Brenig2006d}.

\begin{figure}[t]
\begin{center}\includegraphics[%
  width=0.78\columnwidth,
  keepaspectratio]{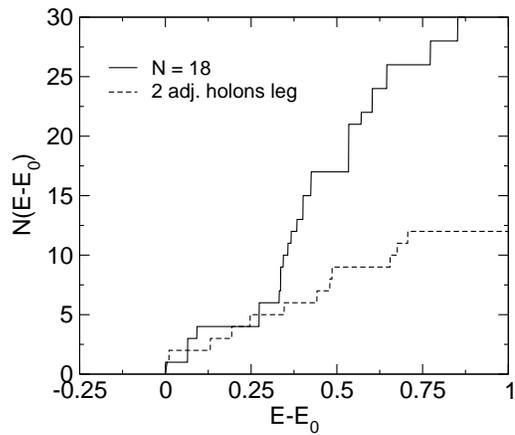}\end{center}
\vspace{-0.5cm}
\caption{\label{fig5}
Number of QD singlets
versus energy at $N=18$, contrasting $0$ holons (solid line) with 
$2$ holons (dashed). Holons are located at sites identical to Fig.
\ref{fig4}. $E_{0}$ refers to ground state energy.
}\vspace{-0.4cm}
\end{figure}

While the QD variational space does not contain triplet excitations, 
it is nevertheless instructive to compare the impact of holons on
the singlet DOS between the ED and QD calculations. In Fig. \ref{fig5}, 
we show the number of singlets above the ground state energy in
the QD case for a system size and a holon placement identical to that
of Fig. \ref{fig4}. First, the overall trend for QDs depicted is
remarkably similar to that in ED. Second, the suppression
sets in at somewhat larger energies, of $E$-$E_0\sim0.3$ for QDs, as compared
to $E$-$E_0\sim0.2$ for ED. Finally, it can be seen that in the energy window
depicted the QD singlets make up for approximately 50\% of the total number of singlets in
ED.%

\subsection{Two-holon Energy\label{sub:Two-holon-Energy}}
\begin{figure}[t]
\begin{center}\includegraphics[%
  width=0.8\columnwidth,
  keepaspectratio]{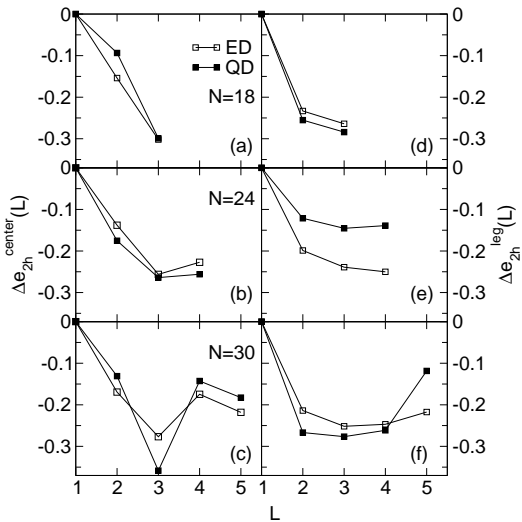}\end{center}
\vspace{-0.3cm}

\caption{\label{fig6}
Comparing two-holon energies at holon separation $L$ from
ED (solid line with open squares) with QDs (solid line with solid
squares) for different systems sizes $N$=18, 24, and 30. (a)-(c)
[(d)-(f)] refer to two holons in [on] the center [leg] of the BTL.}
\vspace{-0.3cm}
\end{figure}

\begin{figure}[t]
\begin{center}\includegraphics[%
  width=0.8\columnwidth,
  keepaspectratio]{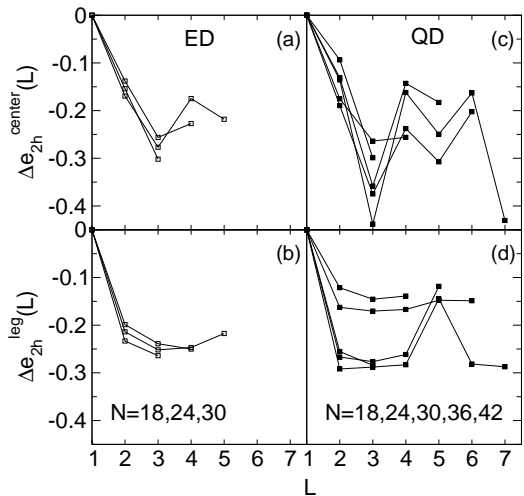}\end{center}
\vspace{-0.3cm}

\caption{\label{fig7}
Evolution of two-holon energies at holon separation $L$
with system size for ED (solid line with open squares)
with $N$=18-30 in (a), (b) and for QDs (solid line with solid
squares) with $N$=18-42 in (c), (d).}
\vspace{-0.3cm}
\end{figure}

Now, we turn to the binding energies
\begin{equation}
\Delta e_{2h}^{a}\left(L\right)=
E_{0}^{a}\left(L\right)-
E_{0}^{a}\left(L=1\right)
\label{eq:2}\end{equation}
of two holons separated by $L$ sites along the center(leg) $=a$
of the BTL, where $E_{0}^{a}$ refers to the corresponding
ground state energies. In Figs. \ref{fig6} and \ref{fig7} we present
two aspects of $\Delta e_{2h}^{a}\left(L\right)$. In the former, we
compare the binding energies obtained from ED with those from QDs
for various systems sizes. First, and remarkably, the ED and QD results
are very similar, both for $a=$ center and leg. Second, these figures
demonstrate that the holons experience a short range repulsive force,
with a maximum binding energy at approximately 3 lattice sites. This
is in sharp contrast to quantum AFMs with VB ground states on bipartite
lattices. There, two holons experience a maximum binding energy if
placed on those nearest-neighbor sites which are occupied by singlet
dimers in the undoped case. Moreover, separating the holons will generate
a string of `misplaced' dimers of length proportional to the holon
separation. This generates a confining potential. Such confinement
cannot be inferred from Fig. \ref{fig6}. Rather, from the largest
system which allows for a direct comparison between ED and QDs, i.e.
$N=30$, an oscillatory behavior at large distances can be anticipated.
This can be corroborated by extending the QD calculation to larger
system sizes as shown in Fig. \ref{fig7}. There, clear oscillations
in $\Delta e_{2h}^{a}\left(L\right)$ can be observed in the right
panel, both on the legs and in the center. In the next section, an
intuitive picture of the two-holon energies in terms of the dimer
coverings will arise. From the preceding, it is tempting to
speculate that mobile holons on the BTL are bound only
weakly and will deconfined already for kinetic energies small
compared to $J$.

Finally, Fig. \ref{fig7} points towards some of the limitations of
the QD approach. While the qualitative behavior is identical for ED
and QDs, i.e. panels (a) vs. (c) and (b) vs. (d), panel (d) shows
a significant variation of the absolute value of the QD binding energies,
depending on whether $N/6$ is even or odd. The latter even/odd effect
is absent in the ED results.%

\subsection{Spin-Spin Correlations\label{sub:Spin-Spin-Correlations}}

\begin{figure}[t]
\begin{center}\includegraphics[%
  clip,
  width=1\columnwidth,
  keepaspectratio,
  angle=-90]{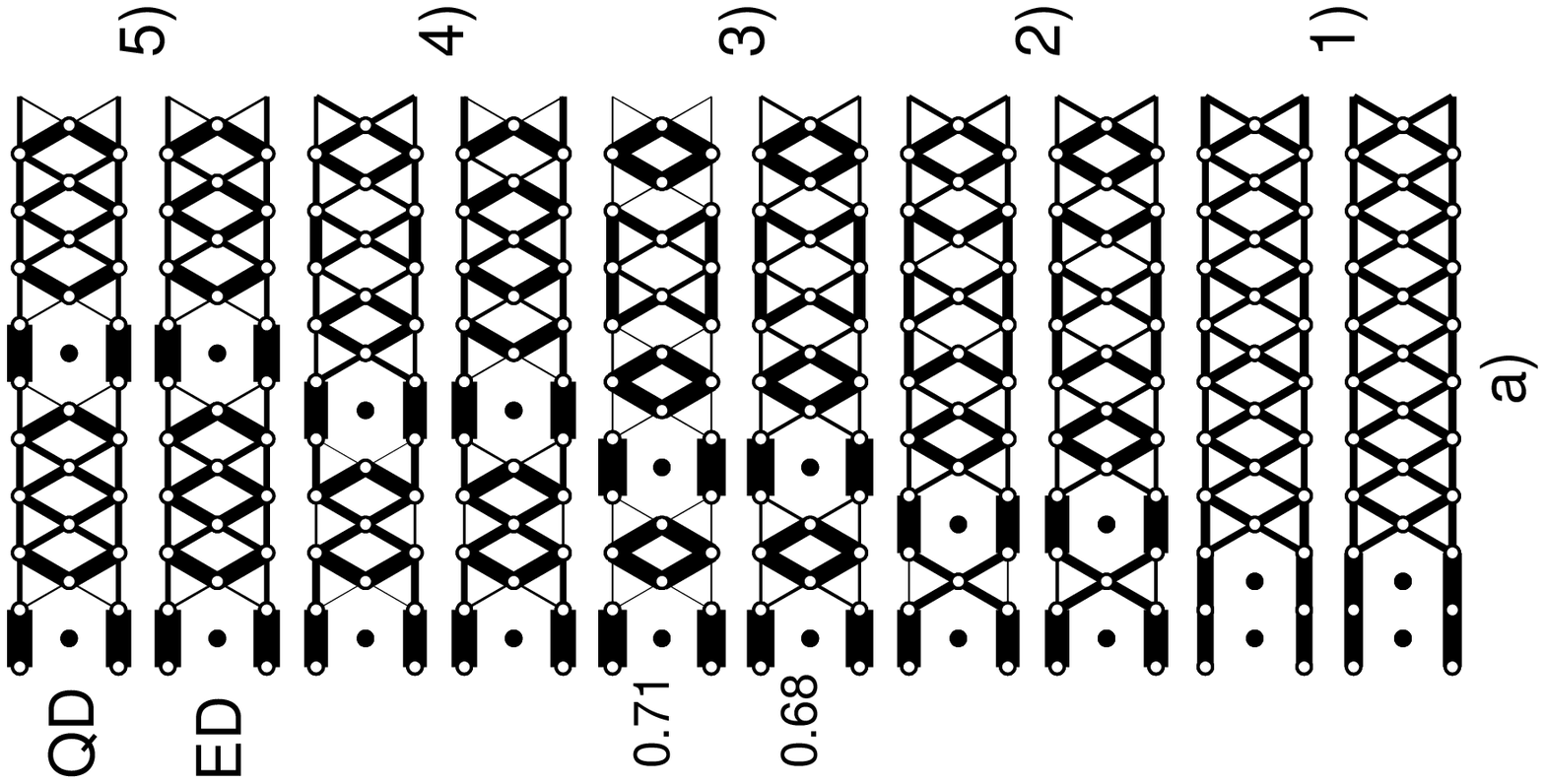}\includegraphics[%
  clip,
  width=1\columnwidth,
  keepaspectratio,
  angle=-90]{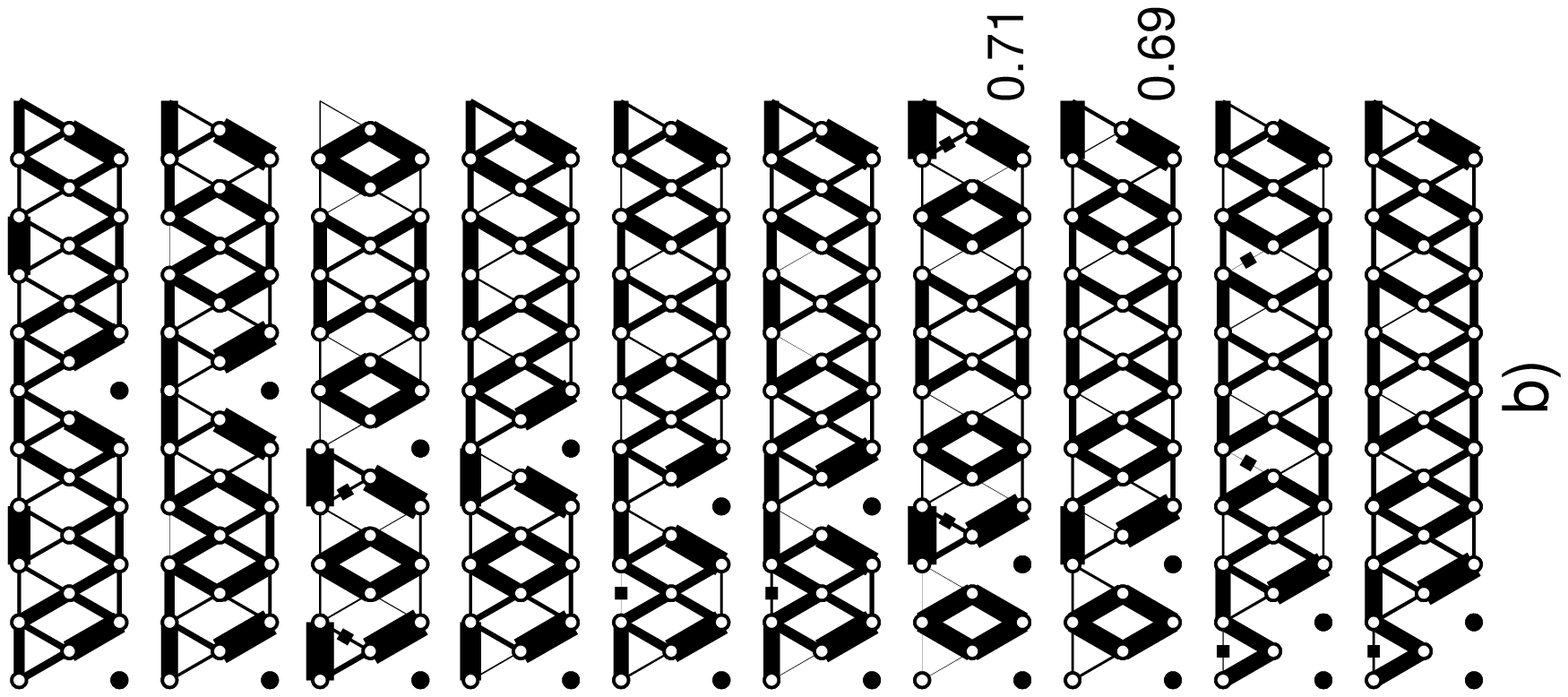}\end{center}
\vspace{-0.5cm}

\caption{\label{fig8}Nearest neighbor spin-spin correlations $s_{lm}$ from
ED and QDs, for two holons (full circles) on the center and leg of
a BTL with $N=30$ for a) and b). The width of solid lines (with vertical
slash) is a linear measure of $-s_{lm}$($s_{lm}$) for $s_{lm}<0$($>0$). 1)$\ldots$5)
labels all non-equivalent two-holon positions. For each 1)$\ldots$5)
the adjacent lower (upper) graph corresponds to one ED (QD) result,
both for a) and b). Numbers refer to maximum values of $-s_{lm}$ observed
for ED and QDs in a) and b).}\vspace{-0.3cm}

\end{figure}

The nearest-neighbor spin-spin correlations 
$s_{lm}=\left\langle \mathbf{S}_{l}\cdot\mathbf{S}_{m}\right\rangle $
are a direct measure of the singlet amplitude on the bond $lm$. Here, 
we consider the impact of two holons on $s_{lm}$ at zero temperature.
Fig. \ref{fig8} summarizes our results, both for ED and QDs on the
largest system for which we have performed ED, i.e. $N$=30. In this
figure $s_{lm}$ is visualized in terms of `bond-thickness'. First,
we note that $s_{lm}$ fulfills the sum rule
\begin{equation}
E_{0}\left(L\right)=\sum_{lm}s_{lm}
\label{eq:3}\end{equation}
where $E_{0}\left(L\right)$ refers to the ground state energy with
two holons in a relative configuration denoted by $L$. Using the
results from Figs. \ref{fig6}, \ref{fig7}, and \ref{fig2}, 
the sum rule has been checked to hold for all cases we have studied,
both for ED and QDs and also for the undoped case, which will not be
discussed here. From Fig. \ref{fig8}, it is evident that the holons
introduce a strong polarization into the magnetic background, which
leads to spatial oscillations of the spin-spin correlations, with
a pattern depending on the separation of the holons. This has to be
contrasted against the undoped case, in which $s_{lm}$ is almost
homogeneous along the BTL with only a very small transverse difference
between the central rungs and the legs. Specifically, the figure demonstrates
that $s_{lm}$ tends to be largest in the vicinity of the holons.
This corroborates the picture of a doping induced reduction of singlet
fluctuations, due to singlet-binding to the holons as discussed in
Sec. \ref{sub:Density-States}. In fact, the maximal values of $|s_{lm}|$
in the vicinity of the holons as shown in the figure are close to
$3/4$, implying a singlet amplitude of $1$ on those bonds. Note 
that for a small number of bonds, ferromagnetic correlations of a rather
small absolute magnitude arise, both in ED and for QDs.

The tendency of the system to accommodate singlets next to the holons
provides for a direct interpretation of the holon-repulsion, namely
the number of `strong' singlets which can be bound to the vicinity of
the holons is less if their separation is small. This can be read
off directly from Fig. \ref{fig8} in going from panel 1) to 5). 

For the majority of holon
placements depicted in Figs. \ref{fig8}a) and b), namely a) 1), 2), 3), and 5) as 
well as b) 1), 2), and 3), the qualitative real-space structure of $s_{lm}$ is 
remarkably similar for ED and QDs. For group
a) 4) there are some differences observable right of the second holon.
In groups b) 4) and 5) there are clear qualitative differences between ED and QDs, 
especially on the leg free of holons. This suggests that QDs provide
for a better description of holons in the `bulk' of the BTL than on its
boundaries.

\begin{figure}[t]
\begin{center}\includegraphics[%
  width=0.8\columnwidth,
  keepaspectratio]{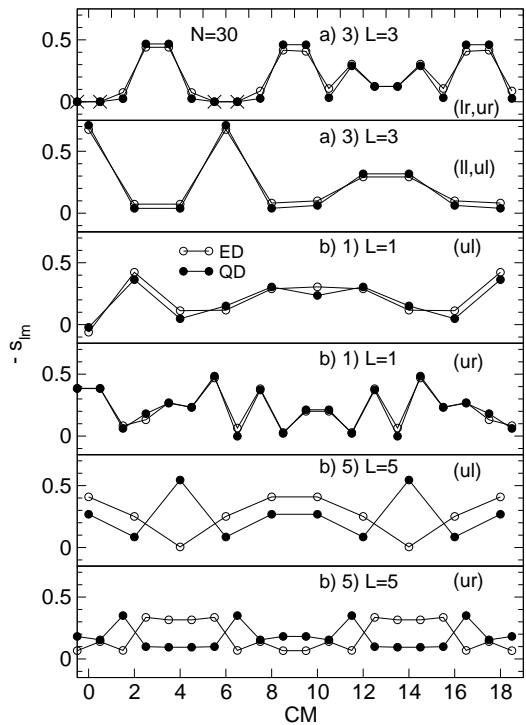}\end{center}
\vspace{-0.3cm}
\caption{\label{fig9}Negative nearest neighbor spin-spin
correlations $-s_{lm}$ along the upper(lower) leg(rung) for u(l)l(r) 
from ED(QDs) for open(closed) circles with two holons and $N$=30, 
corresponding to Fig. \ref{fig8} and panels a) 3), b) 1), and b) 5). 
CM labels `center of mass' of bonds. Crosses mark bonds linked to 
holons where $s_{lm}$ has been set to 0.}\vspace{-0.3cm}

\end{figure}

For a quantitative comparison between ED and QDs, we turn to Fig.
\ref{fig9}, where we show several cuts along the BTL through the legs 
and rungs displaying the values of $-s_{lm}$ versus a measure of distance 
given by a `1D center-of-mass' $CM$ of each bond. $CM$ runs in steps of 
2 (1/2) from 0 (-1/2) to $2N/3-2$ ($2N/3-3/2$) on the legs (rungs). The 
top four panels in Fig. \ref{fig9} are representative of those cases in 
Fig. \ref{fig8} which suggest qualitative agreement between ED and QD. 
It is readily apparent that there is also excellent quantitative agreement. 
The lower two panels in Fig. \ref{fig9} refer to $-s_{lm}$ on the upper 
leg (rung) for the worst case of Fig.
\ref{fig8} b) 5), where ED and QD show strong qualitative differences.

In the following, we provide additional discussion of
the results shown in Figs. \ref{fig6} - \ref{fig9}.
When both holons are positioned in the center axis (see Fig. \ref{fig8}a), reflection
symmetry around this axis is retained. This allows for the formation 
of highly symmetric and simple spin structures. First, as noted above, 
isolated singlets form at the bonds situated closest to the holons 
for all holon separations. Second, more subtle geometrical constraints 
decide if other spin structures (e.g., loops with even numbers of 
spins or short chains) will form, where they will be located, and how much 
singlet correlations there will be between them. For example, in Fig. \ref{fig8}a) 3), 
at a holon distance of $L=3$, apart from the four isolated dimers, one
can clearly notice the presence of 4- and 8-spin loops in the shape of 
three diamonds and a four-pointed star, respectively. These structures are only weakly 
connected by singlet correlations \cite{notea}. Obviously, the distance between the 
holons is one of the controlling parameters in this respect. For example, 
at distance $L=5$, open 3-spin chains are present in the shape of
boomerangs, besides diamonds and dimers. However, in this case, these 
structures are not as independent from each other, i.e., the singlet 
links between them are stronger, as compared to $L=3$.
Therefore, some holon distances are more favorable than others 
in allowing for the formation of such independent `frustration 
releasing' structures. This leads to the oscillating behavior 
of the binding energy in Fig. \ref{fig7}c) \cite{noteb}. Finally, since 
weakly connected dimers, boomerangs, and diamonds can be 
described rather well by the QD basis, the very good quantitative 
agreement between ED and QD for holons located in the center 
axis is no surprise.

When the holons are both on the same leg, reflection symmetry 
around the central axis is lost. Therefore, apart from the dimers 
close to the holons, simple structures as described above are 
less favorable and form only for some holon distances, as for example 
the diamonds seen at $L = 2$ in Fig. \ref{fig8}b). Instead, and
in contrast to the centered holons, more complex correlations 
of the spins occur, which involve the presence of `spin chains' 
of considerable length.
These cannot form when the holons are in the center axis because 
of symmetry considerations and avoidance of uncorrelated free spins.
These chain structures with longer range spin-spin correlations 
are not well described by the QD basis. Their presence leads to 
quantitative discrepancies with the ED results. The difference 
in agreement between Fig. \ref{fig7}(a) and (c) versus Fig.
\ref{fig7}(b) and (d), as well as the discrepancies in the two 
lower two panels in Fig. \ref{fig9} corroborate this interpretation \cite{notec}.

\section{Conclusions}

In summary, we have performed a complementary numerical analysis of a
geometrically frustrated quantum spin ladder with and without static holes
using exact diagonalization and a truncated basis of quantum
dimers. For the undoped system, we have shown that dimers allow for a
reasonable approximation of the ground state energy and the low
temperature specific heat. In the doped case, we have shown that the system can release
frustration through binding of singlets and other extended unfrustrated
spin structures to the static holes. Results for the holon binding energies
and the nearest-neighbor spin correlations confirm this picture. Analysis
of dynamic correlation functions, as e.g. magnetic Raman scattering, and the
finite size scaling of overlaps between ED and QD states, will be
presented elsewhere \cite{next}.

\begin{acknowledgments}
Fruitful discussions with E. Dagotto and A. Albuquerque are gratefully
acknowledged. One of us (W.B.) acknowledges the kind hospitality of the
National High Magnetic Field Laboratory at Florida State University in the
early stages of this project. Part of this work has been supported by DFG
grant No. BR 1084/2-2 and BR 1084/2-3. G.M. acknowledges support from
Research Corporation (Contract No. CC6542). G.M. thanks E. Dagotto for the
use of a 32 Gigabytes Altus 3400 quad-Opteron workstation located at
University of Tennessee in Knoxville. 
\newpage
\end{acknowledgments}


\begin{thebibliography}{35}
\bibitem{notea} The analysis of figures similar to 
Fig.~\ref{fig8} for smaller clusters (not shown)
presents the same picture, i.e., at distance $L=3$ (holons in center axis)
these simple structures are very well defined and quite weakly connected to each other,
resulting in a strong binding energy for holons at this particular distance.
\bibitem{noteb}  It is interesting to note that a 
criterion that appears to be followed in the formation of
these structures is the avoidance of isolated `free' spins.
\bibitem{notec} The presence of these spin-chain structures is easier to spot in 
spin-spin correlations for smaller clusters than the one shown 
in Fig.~\ref{fig8} (with 30 sites). Therefore, the study of varying cluster 
sizes at constant holon doping is needed for a more detailed 
analysis of the properties of these structures.
\end{thebibliography}
\end{document}